\documentclass[twocolumn,amsmath,amssymb,10pt,aps]{revtex4-1} 
\usepackage{graphicx}
\usepackage{latexsym}
\usepackage{amsmath}
\usepackage{amsthm}
\usepackage{amssymb}
\usepackage{epstopdf} 
\usepackage{enumerate}
\usepackage{setspace} 
\usepackage{dcolumn}
\usepackage{bm}
\usepackage{setspace} 
\usepackage{slashed}
\usepackage{color}

\usepackage{tikz}
\usetikzlibrary{arrows,shapes}
\usetikzlibrary{decorations.pathmorphing}	
\usetikzlibrary{decorations.markings}
\tikzset{
    vector/.style={decorate, decoration={snake}, draw},
	provector/.style={decorate, decoration={snake,amplitude=2.5pt}, draw},
	antivector/.style={decorate, decoration={snake,amplitude=-2.5pt}, draw},
    fermion/.style={draw=black, postaction={decorate},
        decoration={markings,mark=at position .55 with {\arrow[draw=black]{latex'}}}},
    ghost/.style={dashed,draw=black, postaction={decorate},
        decoration={markings,mark=at position .55 with {\arrow[draw=black]{>}}}},
    fermionbar/.style={draw=black, postaction={decorate},
        decoration={markings,mark=at position .55 with {\arrow[draw=black]{latex'}}}},
    fermionnoarrow/.style={draw=black},
    gluon/.style={decorate, draw=black,
        decoration={coil,amplitude=4pt, segment length=5pt}},
    scalar/.style={dashed,draw=black, postaction={decorate},
        decoration={markings,mark=at position .55 with {\arrow[draw=black]{>}}}},
    scalarbar/.style={dashed,draw=black, postaction={decorate},
        decoration={markings,mark=at position .55 with {\arrow[draw=black]{<}}}},
    scalarnoarrow/.style={dashed,draw=black},
    electron/.style={draw=black, postaction={decorate},
        decoration={markings,mark=at position .55 with {\arrow[draw=black]{>}}}},
	bigvector/.style={decorate, decoration={snake,amplitude=4pt}, draw},
	flowarrow/.style={postaction={decorate},decoration={markings, mark=at position 0.5 with {\arrow{latex'}}}}}
	
\usetikzlibrary{shapes,snakes,arrows,chains,matrix,positioning,scopes,calc}
\usetikzlibrary{decorations.markings}
\tikzset{/pgf/decoration/.cd,
    number of sines/.initial=10,
    angle step/.initial=20,
}
\newdimen\tmpdimen
\pgfdeclaredecoration{complete sines}{initial}
{
    \state{initial}[
        width=+0pt,
        next state=move,
        persistent precomputation={
            \pgfmathparse{\pgfkeysvalueof{/pgf/decoration/angle step}}%
            \let\anglestep=\pgfmathresult%
            \let\currentangle=\pgfmathresult%
            \pgfmathsetlengthmacro{\pointsperanglestep}%
                {(\pgfdecoratedremainingdistance/\pgfkeysvalueof{/pgf/decoration/number of sines})/360*\anglestep}%
        }] {}
    \state{move}[width=+\pointsperanglestep, next state=draw]{
        \pgfpathmoveto{\pgfpointorigin}
    }
    \state{draw}[width=+\pointsperanglestep, switch if less than=1.25*\pointsperanglestep to final, 
        persistent postcomputation={
        \pgfmathparse{mod(\currentangle+\anglestep, 360)}%
        \let\currentangle=\pgfmathresult%
    }]{%
        \pgfmathsin{+\currentangle}%
        \tmpdimen=\pgfdecorationsegmentamplitude%
        \tmpdimen=\pgfmathresult\tmpdimen%
        \divide\tmpdimen by2\relax%
        \pgfpathlineto{\pgfqpoint{0pt}{\tmpdimen}}%
    }
    \state{final}{
        \ifdim\pgfdecoratedremainingdistance>0pt\relax
            \pgfpathlineto{\pgfpointdecoratedpathlast}
        \fi
   }
}

\begin{document}
\title {Spin-lattice Coupling in U(1) Quantum Spin Liquids}

\author{Sangjin Lee}
\author{Eun-Gook Moon}
\thanks{egmoon@kaist.ac.kr}

\affiliation{Department of Physics, Korea Advanced Institute of Science and Technology, Daejeon 305-701, Korea}

\date{\today}

\begin{abstract}  
Quantum spin liquids (QSLs) are exotic phases with intrinsic massive entanglements. 
Instead of microscopic spins, fractionalized particles and gauge fluctuations are emergent, revealing QSLs' exotic natures. 
{Quantum spins with strong spin-orbit coupling on a pyrchlore lattice, {for example Pr$_2$Zr$_2$O$_7$,} are suggested to host a U(1) QSL with emergent photons, gapless excitations as well as emergent monopoles.  }
One of the key issues in QSLs is an interplay between emergent degrees of freedom of QSLs and conventional degrees of freedom, and we investigate the interplay by constructing a general theory of spin-lattice coupling in U(1) QSLs. 
We find that the coupling induces characteristic interplay between phonons and {photons}. For example, {photons} become qualitatively more stable than phonons at low temperature. 
We also propose mechanisms to detect emergent photons in experiments such as sound attenuation and thermal transport relying on spin-lattice coupling in U(1) QSLs. 
\end{abstract}

\maketitle
\section{introduction}
Strong correlation between localized spins, induced by geometric frustration or quantum fluctuation, is an impetus of QSLs. 
Onsets of conventional order parameters are prohibited by the correlations, and the Landau paradigm of phase transitions become{s} inapplicable. 
Instead, novel concepts and platforms are called for, for example, fractionalized particles, gauge structures, and topological orders, which encodes massive entanglement and non-local characteristics of QSLs \cite{Zhou, Balents, Wen, Sachdev}. 

Mysterious natures of QSLs start to be uncovered by recent advances in experiments. 
Organic materials, for example, {have been suggested to host} fermionic excitations with finite spinon Fermi surfaces \cite{organic1, organic2}. 
Pyrochlore lattices with strong spin-orbit couplings are reported to host exotic phases \cite{Castelnovo, Machida, ashvin, Moon, Kondo, Tokiwa, Kimura, Armitage, Ong, SBLee, Kenzelmann,Noh}. 
Especially, QSLs with U(1) gauge structure are suggested to be realized in magnetic insulating pyrochlores  such as {Pr$_2$Zr$_2$O$_7$ and Yb$_2$Ti$_2$O$_7$}  \cite{Hermele, Shannon, Ross-PRX11,Gingras,gangchen}.
The U(1) gauge structure indicate emergent electrodynamics out of localized moments, which allows exotic collective low energy  excitations such as  emergent photons and monopoles in analogy with quantum electrodynamics in our universe. 
{We stress that emergent photons have gapless energy spectrum and their presence is one of the hallmarks of massive entanglements of localized spins. }
Thus, their identification may be direct evidence of exotic quantum matter.
{However, recent heat capacity measurements, the one of the smoking-gun experiment for U(1) QSLs, have an ambiguity because of nuclear Schottky anomaly \cite{Kimura,Sato}. }

Spin-lattice coupling in QSLs connects exotic excitations from spins and conventional phonons from a lattice, and it is drastically different from one in magnetically ordered phases \cite{Oleg}.  
Microscopic spins are {not useful} concepts in QSLs, and thus {{the conventional spin-lattice coupling Hamiltonian is not a good starting point}.
Instead, it is necessary to find effective spin-lattice couplings in QSLs from {the} scratch, and thus we construct a generic phenomenological theory by employing lattice symmetries and gauge-invariance. 
Using our generic theories, we show that perturbative calculations are valid in {the} deconfined QSLs and find striking consequences of spin-lattice coupling in U(1) QSLs, for example, qualitatively different decay rates of emergent photons and acoustic phonons.
The characteristic behaviors of decay rates {should} appear in physical quantities, and we propose conditions to detect a U(1) QSL with magnetic field dependence. 

\section{Model: spin-lattice coupled Hamiltonian}
 
Let us consider a generic Hamiltonian with spin and lattice degrees of freedom, 
\begin{eqnarray}
H &=& \sum_{ i,j } J^{\alpha \beta}_{ij}(\{x\}) S_{i}^{\alpha} S_{j}^{\beta} -  {{B}^\mu_{ext} \sum_i  g_{B,i}^{\mu \nu}    {S}^\nu_i } +H_L.   
\end{eqnarray}
An exchange interaction $J^{\alpha \beta}_{ij}(\{x\}) $, an external magnetic field $\vec{B}_{ext}$, and Zeeman coupling constant {tensor $g_{B,i}^{\mu \nu}$} are used. 
The indices ($\alpha, \beta)$ are for spin components, and $(i,j)$ are positions of localized spins. 
A position of $S_i^{\alpha}$ is decomposed into an equilibrium position, 
$\{ \vec{R}_i\}$, and a deviation $\{ \vec{x}_i\}$ from $\{ \vec{R}_i\}$.

Setting $J_{ij}^{\alpha \beta} \equiv  J_{ij}^{\alpha \beta} (\{ x=0\})$, a pure spin Hamiltonian is obtained, $H_{S} = \sum_{\langle i,j \rangle} J^{\alpha \beta}_{ij} S_{i}^{\alpha} S_{j}^{\beta} -  {{B}^\mu _{ext} \sum_i g_{B,i}^{\mu \nu}    {S}^\nu _i}$.  
The exchange term ($J_{ij}^{\alpha \beta}$) is constrained by symmetries of the system. 
A lattice Hamiltonian $H_{L} $ consists of a harmonic term and anharmonic interaction terms. 
The harmonic one may be described by
{$H_L^0=  \sum_{k,\lambda} \big[\frac{p_{-k,\lambda}p_{k,\lambda} }{2M} + \frac{M}{2}  \xi_{L, \lambda}(k)^2 x_{-k,\lambda} x_{k, \lambda} \big]  
$}with a mass of ions ($M$). In terms of creation / annihilation operators, the Hamiltonian becomes {$H_L^0 = \sum_{k,\lambda} \xi_{L, \lambda}(k) a_{k, \lambda}^{\dagger} a_{k, \lambda} $ }with a polarization index $\lambda$. The anharmonic one contains higher order creation / annihilation operators,  $H_{L,anh} = O((a,a^{\dagger})^3)$. 
At long-wave length, acoustic phonons are dominant with the dispersion relation, {$\xi_{L,\lambda}(k)=v_{\lambda}|k|$}. A typical value of acoustic-phonon velocities is $v_{\lambda} \sim  10^3 {\rm m/s} $ \cite{Petrenko}.

We notice that the external magnetic field directly couples to spins but not to lattice degrees of freedom.
It is particularly useful to consider the regime where the Zeeman energy is much bigger than spin-exchange but smaller than the Debye frequency ($J_{ij} \ll  \mu_B B_{ext} \ll \omega_D $) with the Bohr magneton, $\mu_B$ { \cite{J1, J2, J3, Kimura,  debye1, debye2, debye3}} . 
In this regime, spin degrees of freedom are fully polarized while phonons are barely affected by the magnetic field. 
Thus, all dynamical quantities are determined by phonons, and thus phonon-only contributions to {the} physical quantities {can} be extracted.

The position dependent exchange interaction is the main source of the spin-lattice coupling, and the Taylor expansion gives,
\begin{eqnarray}
J_{ij}^{\alpha \beta} (\{x\}) - J_{ij}^{\alpha \beta} =  (\vec{x}_{ij} \cdot \nabla_{\vec{r}}) J_{ij}^{\alpha \beta} (\vec{r})|_{\vec{r} = \vec{R}_{ij}}+ O(x^2), \nonumber
\end{eqnarray}
with $\vec{x}_{ij} = \vec{x}_i-\vec{x}_j$ and $\vec{R}_{ij} = \vec{R}_i-\vec{R}_j$. 
Restoring spin operators, the first-order term becomes,
\begin{eqnarray}
H_{L-S} = \sum_{i,j} (\vec{x}_{ij} \cdot \nabla_{\vec{r}}) J_{ij}^{\alpha \beta} (\vec{r})|_{\vec{r} = {\vec{R}_{ij}}} S_i^{\alpha} S_j^{\beta}, 
\end{eqnarray}
describing a process between one-phonon and two spin excitations. It is straightforward to consider higher order terms. 
In a magnetically ordered state, one can replace $S_{i}^{\alpha} \rightarrow \langle S_{i}^{\alpha} \rangle + \delta S_i^{\alpha}$, and the fluctuations can be treated in the standard way, for example, by using the Primakoff-Holstein representation\cite{Kreisel}. 

In QSLs, however, the standard approach fails because mean values vanish. Instead, one needs to construct an {effective Hamiltonian} between emergent degrees of freedom and phonons. 
Two criteria are used; gauge invariance and lattice symmetries. 
The gauge invariance {\it must} be respected since physical observables are gauge independent. Gauge invariant operators include  emergent electric / magnetic  fields ($\vec{e}, \vec{b}$), densities ($\rho_e, \rho_b$), and currents ($\vec{j}_e, \vec{j}_b$) of emergent electric / magnetic monopoles. Lattice symmetries assign quantum numbers to gauge invariant operators. For example, magnetic monopole density breaks inversion and time reversal symmetry (see Table I). 
Lattice symmetries {enforce} strain tensor as a key ingredient of lattice degrees of freedom. Acoustic phonons are dominant at low energy, and all couplings of acoustic phonons appear with strain tensor because acoustic phonons are  Goldstone bosons.

To be specific, let us consider a system with cubic, time-reversal (TRS), and inversion (Inv) {symmetries}, motivated by pyrochlore systems. The presence of the triplet representation ($T$) of the cubic group is particularly useful {because} it can be treated as  a `vector' of the cubic group. 
The { {\it strain tensor}, {\it {$ \mathcal{P}_{mn} \equiv \frac{1}{2} (\partial_{m} x_{n} + \partial_{n} x_{m} )$},}} contains the two indices ($m$,$n$) of the $T$ representations, and one can use the relation, 
$T \otimes T \otimes T \otimes T = 1 \oplus +\cdots$ to find the trivial representation.
By using the fact that $\vec{e}, \vec{b}$ are in the $T$ representation, we find one coupling term, 
\begin{eqnarray}
H_{L-\nu} = \int_x \mathcal{P}_{mn} \big[g_b {c^2  b_{m}  b_{n} + g_e e_{m}  e_{n}} \big]. 
\end{eqnarray} 
{ As in the low temperature limit, the Umklapp process is naturally suppressed, we are going to focus on the normal process and we adopt the continuum notation for simplicity, hereafter. Note that it is easy to incorporate the Umklapp process in the low energy effective theory level as discussed in previous literatures \cite{Fradkin}.     }
Each term describes a scattering between two-photons and one-phonon.
Note that  $\int_x \mathcal{P}_{mn} ( \partial_{m} b_{n} +  \partial_{m} b_{n})$ is forbidden by TRS, and $\int_x \mathcal{P}_{mn} ( \partial_{m} e_{n} +  \partial_{m} e_{n})$ is { suppressed} by energy-momentum conservation (see below).
 
The same principles of gauge-invariance and lattice symmetries {apply} to all other physical operators. With emergent {electric monopole}, $E$-particles, the coupling term is 
 \begin{eqnarray}
 H_{L-e} = \int_x \mathcal{P}_{mn} \big[  j^{m}_{e}  j^{n}_{e} + \partial^{m} \rho_e \partial^{n} \rho_e +\cdots \big], \nonumber
 \end{eqnarray}
which describes four electric monopoles and one phonon process. 
 Coupling constants in LHS terms are implicit and, similarly, one can construct the coupling between acoustic phonons and {magnetic monopole}, M-particles, $H_{L-b}$, as well as higher order terms.  
Remark that our construction can be easily extended to systems with other symmetry groups.
 
The external magnetic field effects in QSLs can be similarly obtained.  
Since the external field does not directly couple to {the} lattice degrees of freedom, one can focus on couplings to emergent degrees of freedom of spins. 
The external magnetic field is also in the $T$ representation breaking TRS, and it {only couples} to emergent magnetic field,  
{$H_{Z} = -  \int_x  g_{Z}^{\mu \nu} {b}^\mu (x) {B}^\nu _{ext} $}.
It is obvious that a uniform external magnetic field induces a uniform emergent magnetic field $\vec{b}_0 \neq0$. The uniform field does not change the dynamics of photons, but it affects $E$-particles' dynamics significantly similar to the case of electrons under magnetic fields. 
Thus, it is important to keep $E$ particle dynamics in the discussion of external magnetic field dependence, and  
the energy gap of $E$ particles shifts under $\vec{B}_{ext}$ as in the Landau level of electrons, giving 
{$\Delta_E (B_{ext}) = \Delta_E +  |g_{Z}^{\mu \nu} {B}^\mu_{ext}|$ { in the weak magnetic field limit}.}
The coupling constant {$g_Z^{\mu \nu}$} is related with the mass and charge of $E$-particles.

\begin{table}[tb!]
\begin{tabular} {|c|c|c|c|c|c|c|}
\hline 
 & ~~$\rho_E $~~ & ~~$\rho_M $~~  & ~~ $\vec{\varepsilon}~~ $   & ~~$\vec{b}$~~  & ~~$\vec{j}_E $~~  & ~~$\vec{j}_M $~~  \\ \hline \hline
TRS & $+$ & $-$ & $+$ & $-$ & $-$ & $+$ \\
 \hline  
INV & $+$ & $-$ & $-$ & $+$ & $-$ & $+$ \\ \hline 
\hline
\end{tabular}
\caption{
Symmetry properties under time-reversal symmetry (TRS) and inversion (INV). 
}
\end{table}

We first consider stability of the QSLs under the spin-lattice coupling. 
Monopole excitations in U(1) QSLs are gapped, so one can focus on acoustic phonons and emergent photons for the stability of U(1) QSLs. Moreover, we ignore anharmonic interactions of phonons and compactness of U(1) gauge fields for now 
to see intrinsic effects of spin-lattice coupling and check its self-consistency later.  

The effective Hamiltonian of acoustic phonons and emergent photons is  
\begin{eqnarray}
H_{eff} =   \int_x \big(\frac{\vec{b}\,^2 }{2 \mu_0}  +  \frac{\epsilon_0  \vec{\varepsilon}\,^2 }{2} \big)  + H_L+ H_{L-\nu}
\end{eqnarray}
with emergent permittivity and permeability $\epsilon_0, \mu_0$. 
Linear dispersion relations of the two excitations, $\xi_{L} (k)=v |k|$ and $\xi_{\nu}(q) = c|q|$, allow a scale analysis with $[x] =-1$ and a dynamical critical exponent. Imposing $z=[H_{eff}]=1$, we find 
\begin{eqnarray}
[\vec{b}] = [\vec{\varepsilon}]  =  [\mathcal{P}_{ab}]= 2,\quad \rightarrow \quad [g_{e,b}]=-2, \nonumber
\end{eqnarray} 
and it is clear that the coupling constants are irrelevant in the renormalization group sense. 
Therefore, U(1) QSLs is stable under spin-lattice coupling, and their properties can be understood by perturbative calculations.  

The stability analysis allows us to incorporate effects of the anharmonic terms and compactness of U(1) QSLs. 
The anharmonic term is associated with a cubic term such as $x_{a} x_{b} x_{c}$. Note that gauge invariance and lattice symmetries prohibit interaction terms with an odd number of the emergent electromagnetic fields. 
Simple power counting shows that the two operators are irrelevant, and the perturbative analysis should be valid.

 \section{decay rate of photons and phonons}
Next, we calculate decay rates of the two excitations, emergent photons and acoustic phonons under the scattering mechanism of $H_{L-\nu}$. 
Even though the two excitations have the linear dispersion relation at long wave length, 
their velocities are fundamentally different  because a phonon velocity is associated with Debye frequency ($\omega_D$) and one of emergent photons is associated with typical spin-exchange interactions which are much smaller than the Debye frequency. 
For example, pyrochlore structures have typical phonon velocity {$v_L \sim 10^3$m/s} and their emergent photon velocities are estimated to be $c \sim 10$m/s. Therefore, a dimensionless parameter $\alpha \equiv \frac{v}{c} \gg1$ is naturally introduced. Note that the non-unity velocity ratio {significantly constrains phase spaces of phonon-photon scattering}, which makes the coupling term, $\int_x \mathcal{P}_{mn} ( \partial_{m} e_{n} +  \partial_{n} e_{m} )$, {disallowed because $\xi_L(k) = \xi_{\nu}(k)$ works only at $k=0$ point which is negligible in the phase space, }{ as in the optical experiment where speed of light is highly mismatched to velocities of elementary excitations}. 

The {only} allowed scattering process between two-photons and one-phonon from $H_{L-\nu}$ is
\begin{eqnarray}
(\vec{k}, \lambda'' ; L  ) \Leftrightarrow (\vec{q}, \lambda ; \nu) + (\vec{q}\,', \lambda' ; \nu) \nonumber
\end{eqnarray}
constrained by energy and momentum conservations,
\begin{eqnarray}
\xi_{L, \lambda''}(\vec{k}) = \xi_{\nu, \lambda}(\vec{q}) + \xi_{\nu,\lambda'} (\vec{q}\,'), \quad \vec{q} +\vec{q}\,' = \vec{k} {\;\;  (\text{mod } \vec{G}) }\nonumber
\end{eqnarray}
A reciprocal lattice vector $\vec{G}$, polarizations ($\lambda, \lambda', \lambda''$) of the photons ($\nu$) and phonon ($L$) are explicitly shown. In the normal process ($\vec{G}=0$), we find that $|\vec{k}| \ll | \vec{q}| $ because of $\alpha \gg 1$. 
Precise values of the velocity ratio $\alpha$ depends on polarizations, but one can use the condition  $\alpha \gg 1$ for all polarizations.
It is easy to show that other processes, for example 
\begin{eqnarray}
(\vec{q}, \lambda ; \nu) + (\vec{k}, \lambda' ; L) \Leftrightarrow (\vec{q}+\vec{k}, \lambda'' ; \nu ), \quad {\rm (forbidden)}\nonumber
\end{eqnarray}
is forbidden because of the condition ($ \xi_{\nu} (\vec{q})+\xi_{L}(\vec{k}) \neq \xi_{\nu}(\vec{q}+\vec{k})  $).  
Therefore,  the scattering process only appears when the two photons move almost oppositely.
 
\begin{figure}
\includegraphics[width=3.in]{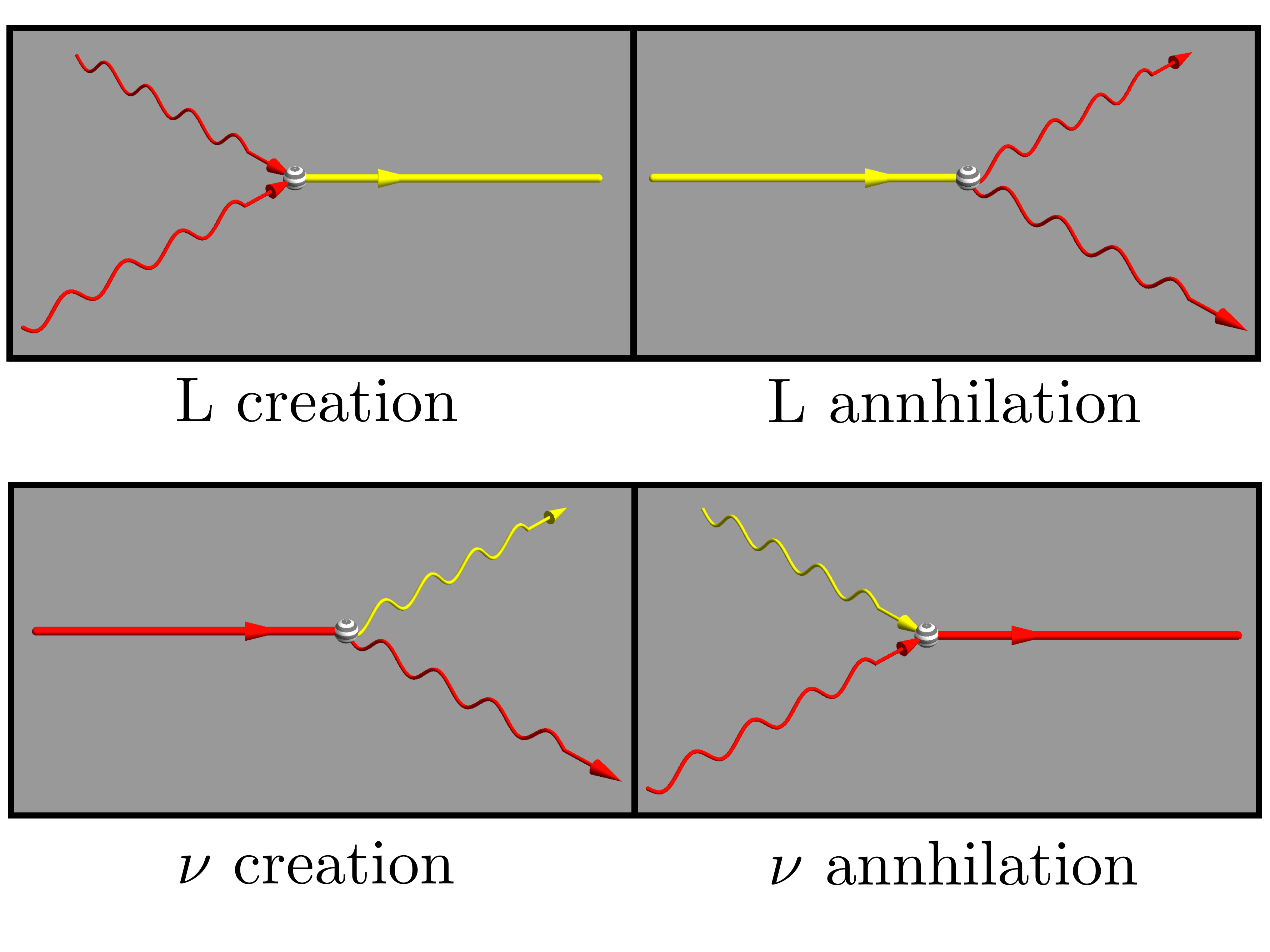}
\caption{ Characteristic scattering mechanisms between phonons and emergent photons under spin-lattice coupling. Wavy lines are for emergent photons ($\nu$) and straight lines are for acoustic phonons (L). Gauge invariance and lattice symmetries prohibit other channels.  }
\end{figure}

 Decay rates of the two excitations are calculated by using the Fermi-Golden rule,  
 $\mathcal{P}_{i \rightarrow f} = \frac{2\pi}{\hbar} |\langle f | H_{L-\nu}| i \rangle|^2 \,\, \delta[\xi]$,
 between the initial state, $|i\rangle = |( \vec{k}, \lambda'' ; L ) \rangle$, and the final state, $|f\rangle = |(\vec{q}, \lambda ; \nu) + (\vec{q}\,', \lambda' ; \nu)\rangle$. The schematic delta function is for the energy conservation, ($\delta[\xi] \equiv \delta(\xi_{L, \lambda''}(\vec{k}) = \xi_{\nu, \lambda}(\vec{q}) + \xi_{\nu,\lambda'} (\vec{q}\,'))$).  
Hereafter, the polarization indices are implicit. 

We find the phonon decay rate,
\begin{eqnarray}
\frac{1}{\tau_{L}(k)} =  \int_q |M|^2 \delta([\xi]) (1+n_B(\xi_{\nu}(\vec{q}))+n_B(\xi_{\nu} (\vec{k}-\vec{q}))),\nonumber
\end{eqnarray}  
and the photon decay rate, 
\begin{eqnarray}
\frac{1}{\tau_{\nu}(q)} =  \int_k |M|^2 \delta([\xi]) (n_B(\xi_{\nu} (\vec{k}-\vec{q}))-n_B(\xi_{L}(\vec{k}))).\nonumber
\end{eqnarray}  
{ The distribution factors in above expressions are obtained by the difference between the creation and the annihilation process whose distribution factors are $n_{B}(\xi_{\nu})$, $n_{B}(\xi_{L})$ for photons and phonons respectively.
For the photon decay rates, distribution factors are determined by
\begin{align*}
&n_{B} (\xi_{\nu}) (1+ n_{B}(\xi_{L})) - (1+n_{B} (\xi_\nu))n_B(\xi_\nu)\\
&\quad = n_{B}(\xi_{\nu}) - n_{B}(\xi_{L}),
\end{align*} 
where the first (second) term for the creation (annihilation) processs of the photon.
Similarly, distribution factor in phonon decay rates can be understood,
\begin{align*}
&(1+ n_{B}(\xi_{\nu,1}))(1+ n_{B}(\xi_{\nu,2}))- n_{B}(\xi_{\nu,1})n_{B}(\xi_{\nu,2}) \\
&\quad = 1+  n_{B}(\xi_{\nu,1})+ n_{B}(\xi_{\nu,2}) .
\end{align*}
}
The matrix elements ($|M|$) is obtained by the Fourier transformation of $H_{L-\nu}$ whose scaling dimension is $[M] =3/2$, determined by the scaling analysis of $H_{L-\nu}$.
Notice that the phase space factors in the two decay rates are {{\it qualitatively different}. } 
The phase space factor of phonons ($1+n_B(\xi_{1})+n_B(\xi_2)$) is from the difference between the decay process with a factor $(1+n_B(\xi_1))(1+n_B(\xi_2))$ and the creation process with a factor $n_B(\xi_1)n_B(\xi_2)$ as shown in Fig. 1(a).   
The one of the photons is obtained by the difference between the decay process with a factor $ n_B(\xi_{1})(1+n_B(\xi_2))$ and the creation process with a factor $(1+n_B(\xi_1))n_B(\xi_2)$ as shown in Fig. 1(b).

Note that the decay rates can be equivalently obtained by evaluating the imaginary parts of the self-energies of photons and phonons (see appendix).
We find the decay rates with two dimensionless functions ($f_{L}$ and $f_{\nu}$), 
{
\begin{eqnarray}
\frac{1}{\tau_{L}(k)} &=& {c g_e^2}\left(\frac{T}{c}\right)^5 f_L\left( \frac{\xi_L(k)}{T}, \alpha; g_b, g_e \right)  \nonumber \\
\frac{1}{\tau_{\nu}(q)} &=&  {c g_e^2}\left(\frac{T}{c}\right)^5 f_{\nu}\left( \frac{\xi_\nu(q)}{T}, \alpha; g_b, g_e \right).  \nonumber
\end{eqnarray} 

Asymptotic behaviors of the dimensionless functions are as follows. The phonon function shows $f_{L} (x) \propto \alpha^2  x^4$ for $x \ll 1$, { $f_{L}(x) \propto \alpha^3 x^5$ for $x\gg1$}, and the photon function shows 
$f_{\nu} (x) \propto \alpha^{-5}  x^4$ for $x \ll 1$ and $f_{\nu} (x) \propto \alpha^{-4}  x^5 e^{-x}$ for $x \gg 1$. { In the low temperature limit, decay rate shows quantitatively different behaviors.} Explicit forms of the two functions are presented in the appendix. } 

The decay rates of phonons and photons show qualitative differences. In the limit of $T \ll \xi_L(k), \xi_{\nu}(q)$,
the phonon decay rate is bigger than the temperature independent term, $\tau_{L}^{-1} > c_{0} |k|^5 $ with a positive constant $c_0$ {because of suppression of exponential terms in $1+n_B(\xi_{\nu,1})+n_B(\xi_{\nu,2})$} while the photon decay rate shows $\tau_{\nu}^{-1} \propto e^{-\frac{\xi_{\nu}(q)}{T}} $. 
Such qualitatively different decay rates are originated from different phase spaces. 
Namely, the phonon decay rate has the phase space, $1+n_B(\xi_1) +n_B(\xi_2)$, and it is obvious  the temperature independent term dominate over the other terms at low temperature.  
On the other hand, the phase space of the photon decay rate has the factor ($n_B(\xi_1) - n_B(\xi_2)$), which always contain the Boltzmann factor, $e^{-\frac{\xi_{\nu}(q)}{T}}$.
Therefore, at low temperature, the photons are qualitatively more stable than the phonons. 
Accordingly, the mean free path of the photon ($l_{\nu}(q) \equiv c \tau_{\nu}(q)$) is much longer than one of the phonons ($l_{L}(q) \equiv v \tau_{L}(q)$) at low temperature, 
$l_{\nu}(q) \gg l_{L}(q)$, under the spin-lattice coupling.
 
We emphasize that external magnetic field dependences of the decay rates are also characteristic. 
At low temperature ($T \ll \Delta_E$), one can treat $E$-particles semi-classically with the Maxwell distribution function. The semi-classical analysis with the Mattiessen's rule may be applied and the phonon decay rate receives corrections from thermal excitations of $E$-particles
\begin{eqnarray}
\frac{1}{\tau_{L}(k)}  > g_{e,{b}}^2 |k|^5 + h_{L}(k, T) e^{- \frac{2\Delta_E({B}_{ext})}{T}}, \nonumber
\end{eqnarray}
while the photon decay rate becomes
\begin{eqnarray}
\frac{1}{\tau_{\nu}(q)}  \propto g_{e,{b}}^2 |q|^5 e^{-\frac{\xi_{\nu}(q)}{T}} + h_{\nu}(q, T) e^{-\frac{\Delta_E({B}_{ext})}{T}}. \nonumber
\end{eqnarray}
Explicit forms of the well-defined functions, $h_{\nu,L}$, are presented in the appendix. 
At low temperature, it is obvious that emergent photons are much more sensitive than phonons. 
In the photon decay rate, the spin-lattice and external magnetic fields contribute with the Boltzmann factors, and thus photons with $\xi_{\nu}(q) \sim \Delta_{E}(\vec{B}_{ext})$ are significantly affected. Remarkably, {\it the photon decay rate decreases increasing $\vec{B}_{ext}$} because the excitation energy of $E$-particles increases.
On the other hand, the decay rate of the {phonons} is barely affected by external magnetic field because the dominant channel is the $|k|^5$ term from the spin-lattice couplings as well as anharmonic interaction terms. Such weak $\vec{B}_{ext}$ dependence of the phonons is originated from the absence of the Zeeman coupling between lattices and external magnetic field.

\section{discussion and conclusion}
Our decay rate calculations may be applied to  thermal conductivity ($\kappa = \frac{1}{3} \sum_q c_q v_q^2 \tau_q$) with external magnetic field at low enough temperature where acoustic phonons and the emergent photons become important. 
Our results suggest that main carriers of thermal conductivity are photons because {mean free path of photon} becomes much larger than {that} of phonons in addition to their larger specific heats. This can be tested by magnetic field dependences which would show  magneto-thermal resistance from the decreased photon decay rate under the external magnetic field. 
More detailed thermal transport calculation will be discussed in other places (see also appendix).

Furthermore, the characteristic decay rate of the acoustic phonons in U(1) QSLs may be directly measured by sound attenuation experiments {which may be complimentary to the heat capacity measurement}{ \cite{sndatt1,sndatt3}}. Note that previous literature studied sound attenuation with fermionic spinons with strongly renormalized gauge fluctuations in two spatial dimensions \cite{Lee1,Lee2} in sharp contrast to our discussion. 
It is important to consider not only spin-lattice coupling but also anharmonic interaction between phonons. The Mattiessen's rule give $\tau_{L,tot} ^{-1} (\omega) =\tau_{L,S}^{-1} (\omega)+ \tau_{L,L}^{-1} (\omega)$. The former ($\tau_{L,S}^{-1} $) is from spin-lattice coupling and the latter ($\tau_{L,L}^{-1} $) is from anharmonic interactions {which behaves} as $\tau_{L,L}^{-1} (\omega) \propto \omega T^4$ in the limit of $\omega \ll T$.
In the low temperature limit ($T \ll J$) of U(1) QSLs {\cite{Kimura, J1, J2}} and { frequency within ultrasound range, $\mathcal{O}(10^2 \text{MHz}\sim 0.01K)$}, only emergent photons are important among spin degrees of freedom, and thus our result of {$\tau^{-1}_{L}(\omega) \propto  T \omega^4$ }can be directly used for $\tau^{-1}_{L,S}$. 
Then, we propose that the sound attenuation from photons can be obtained by subtracting sound attenuations with high magnetic field and without magnetic field, 
\begin{eqnarray}
\frac{1}{\tau_{L,tot}(\omega; B_{ext}=0)} -\frac{1}{\tau_{L,tot}(\omega; B_{ext}\rightarrow \infty)} \propto T \omega^4, \nonumber  
\end{eqnarray}
{ which can be understood by the power counting of phonon self-energy calculations from photons: $\omega^5$ from matrix element and integral measure and distribution factor gives $\frac{T}{\omega}$. } 
The linear $T$ dependence may be {the} evidence of the U(1) QSLs.

In conclusion, we present a general theory of spin-lattice coupling and find characteristic interplay between phonons and emergent photons. 
We show that U(1) QSLs are stable under spin-lattice coupling and calculate characteristic decay rates of phonons and photons. 
It is shown that emergent photons are qualitatively more stable than phonons at low temperature. We also propose mechanisms to detect emergent photons in experiments such as sound attenuation and thermal transport.

\section*{Acknowledgement }
 It is our great pleasure to have invaluable discussion with  L. Balents, A. Furusaki, Y. Matsuda, Y. Motome, T. Shibauchi, and Y. Tokiwa. 
 We are indebted to L. Balents for various inputs at the early stage of this work. 
 EGM is particularly grateful to Y. Motome and A. Furusaki for their hospitalities during the visits to the University of Tokyo and RIKEN.  
This work was supported by the POSCO Science Fellowship of POSCO TJ Park Foundation and NRF of Korea under Grant No. 2017R1C1B2009176.

 

\pagebreak
\widetext
\setcounter{equation}{0}
\setcounter{table}{0}
\setcounter{page}{1}
\setcounter{figure}{0}
\setcounter{section}{0}

\makeatletter
\makeatletter
\renewcommand{\theequation}{A\arabic{equation}}
\renewcommand{\thefigure}{A\arabic{figure}}

\begin{center} 
\textbf{Appendix A: Decay rates }
\end{center}
\subsection{Matrix elements}

The phonon-photon coupling is 
\begin{eqnarray}
H_{L-\nu} = \int_x \mathcal{P}_{ab}\big[g_b c^2  b_{a} b_{b} + g_e \varepsilon_{a} \varepsilon_{b} \big].
\end{eqnarray}
In terms of normal modes, the electric / magnetic fields are
\begin{eqnarray}
\vec{b} (x) &\propto& \sum_{q,\lambda} \frac{1}{\sqrt{\xi_{\nu{\lambda}}(\vec{q}) }}  a_{q\lambda} (\vec{q} \times  \vec{\epsilon}_{\lambda}(q)) e^{i q x} + h.c. \nonumber \\
\vec{\varepsilon} (x) &\propto& \sum_{q,\lambda}  \frac{1}{\sqrt{\xi_{\nu{\lambda}}(\vec{q})}}  a_{q\lambda} ( \omega_{q} \vec{\epsilon}_{\lambda}(q)) e^{i q x} + h.c., \nonumber
\end{eqnarray}
and the strain tensor is
\begin{eqnarray}
\partial_{\alpha} x_{\beta} \propto k_{\alpha}  \sum_{k, \lambda} \frac{1}{\sqrt{\xi_{L{\lambda}}(\vec{k}) }} a_{k\lambda} (  {\epsilon}_{\beta \lambda}(k)) e^{i k x} + h.c.
\end{eqnarray}
with $\vec{\epsilon}$ represents polarization vector of photon(phonon).
The coupling term becomes
\begin{eqnarray}
H_{L-\nu} \sim \int_x \sum_{k,q,q'} \frac{q q' k}{\sqrt{\xi_{\nu{\lambda}}(\vec{q}) \xi_{\nu{\lambda'}}(\vec{q'}) \xi_{L{\lambda''}}(\vec{k})   }} (a_{q\lambda}+a_{q\lambda} ^{\dagger})(a_{q'\lambda'} +a_{q'\lambda'} ^{\dagger})(a_{k\lambda''}+a_{k\lambda''}^{\dagger}). \nonumber
\end{eqnarray}

The matrix element is
\begin{eqnarray}
M \sim \frac{q q' k}{\sqrt{q q' k}} \rightarrow [|M|]=\frac{3}{2}.
\end{eqnarray}

\subsection{Phase space} 
The different velocities makes characteristic $\alpha$ dependences of phase space of photons and phonons.
Let us first consider the phase space of phonons, 
\begin{eqnarray}
&&\int _q  \delta\left( (\xi_{L}(\vec{k})- \xi_{\nu} (\vec{q}))^2 - \xi_{\nu}(\vec{k}-\vec{q})^2\right) \nonumber \\
&=& \int _q \,\, \delta\left(v^2 k^2 - c^2 k^2 - 2 v c k q + 2 c ^2 k q \cos(\theta)\right) \nonumber \\
&=& \int_{k (\alpha-1)/2}^{k (\alpha+1)/2} dq \frac{q}{2 c^2 k } = \frac{k}{4c^2} \alpha.
\end{eqnarray}
On the other hand, the photon phase space is 
\begin{eqnarray}
&& \int_k  \delta\left( (\xi_{L}(\vec{k})- \xi_{\nu} (\vec{q}))^2 -( \xi_{\nu}(\vec{k}-\vec{q}))^2\right) \nonumber \\
&=& \int_k  \,\, \delta\left(v^2 k^2 - c^2 k^2 - 2 v c k q + 2 c^2 k q \cos(\theta)\right) \nonumber \\
&=& \int_{2q /(\alpha+1)}^{2q /(\alpha-1)} dk \frac{k}{2 c^2 q } = \frac{q}{4 c^2 } \left(\frac{4}{(\alpha-1)^2}-\frac{4}{(\alpha+1)^2} \right) \nonumber \\
&=& \frac{q}{4 c^2} \frac{16 \alpha}{(\alpha^2-1)^2}.
\end{eqnarray}
This calculation indicates that with a same momentum, phonon decay phase space is much wider than one of photons.

\subsection{Self-energy calculation}
\subsubsection{Photon decay rate}
With the dispersion relations, $\xi_{\nu} (\vec{q})  = c q$, $\xi_{L}(\vec{k})  = v k$, the photon self-energy from phonon is 
\begin{align*}
&\begin{tikzpicture}
\draw[vector,thick](-1.5,0)--(1.5,0);
\draw[dotted,thick](-.9,0)arc(180:0:1);
\end{tikzpicture}\nonumber 
\end{align*}
\begin{align*}
&\rightarrow \Pi _{L\rightarrow \nu} = T\sum_{k_n}\int_k \frac{I_1 } {\left((i k_n)^2 - \xi_{L }(\vec k) ^2\right)\left(( i\Omega - i k_n )^2 - \xi_{\nu}{( \vec{k}-\vec{q})}^2\right)} ,
\end{align*}
where 
\begin{align*}
12I_1  =& \frac{1}{i \Omega}\left((-k_i \epsilon_{L j} ^{*}- k_j \epsilon_{L i} ^{*}) g_e (i \Omega) \epsilon_{\nu q i}  (-i \Omega+ i k_n ) \epsilon_{\nu q-k j }^{ *} (k_a \epsilon_{L,b}+ k_b \epsilon_{L a} ) g_e (-i \Omega) \epsilon_{\nu q a}^{*}  (i \Omega- i k_n ) \epsilon_{\nu q-k b } \nonumber  \right. \\
&+ \left.(-k_i \epsilon_{L j} ^{*}- k_j \epsilon_{L i} ^{*}) g_b c ^2   (\vec{q}\times \vec{\epsilon}_{\nu q})_i (-(\vec{q}-\vec{k})\times \vec{\epsilon}_{\nu q-k }^*)_j   (k_a \epsilon_{L b} + k_b \epsilon_{L a} ) g_b c ^2   (-\vec{q}\times \vec{\epsilon}_{\nu q  }^*)_a ((\vec{q}-\vec{k})\times \vec{\epsilon}_{\nu q-k })_b\right).
\end{align*}
The polarization indices ($\epsilon_{Li}, \epsilon_{\nu}$) are introduced.  We are mainly interested in physics independent of polarizations, so we average over all polarizations.  

We approximate that velocity of longitudinal phonon modes is same as the velocities of transverse phonon mode for simplicity of calculation. Qualitative results of our calculations, for example powers of temperature dependences, are correct though quantitative results would be affected by the approximation.  We use sum-rule for phonon, $\sum_{L.pol} \epsilon_{Li}\epsilon_{Lj}^*= \delta_{ij}$ where the summation is over the transverse and the longitudinal mode, and for photon, $\sum_{\nu.pol} \epsilon_{\nu q i} \epsilon_{\nu q j}^* = \delta_{ij} - \frac{q_i  q_j}{q^2} = P_{ij}(\vec{q})= P_{ji}(\vec{q}) $ where the summation is over the transverse modes,
\begin{align*}
&\Pi_{L\rightarrow \nu}  = \\
&\int_k\;   \frac{g_e^2 (i\Omega ) }{12}  \left(2  k_i k_aP_{ia}(\vec{q}) +2 k_a k_j P_{ia}(\vec{q})P_{ji}(\vec{q}-\vec{k})   + 2    k_j k_b P_{jb}(\vec{q}-\vec{k})                                           \right)  T \sum_{k_n} \frac{-(i\Omega -ik_n)^2}{((i k_n )^2- \xi_{L}(\vec{k})^2)((i \Omega -i k_n)^2 - \xi_{\nu}(\vec{k}-\vec{q})^2)} \nonumber \\
&+ \int_k \; \frac{g_b^2 c^4}{12}  \left((\vec{k}\times\vec{q})^2(2q^2+ 2(\vec{q}-\vec{k})^2)    + 2 (\vec{q}\times(\vec{k}\times\vec{q}))\cdot ((\vec{q}-\vec{k})\times (\vec{k}\times\vec{q}))           \right)  T \sum_{k_n} \frac{-1}{i\Omega  ((i k_n )^2 -\xi_{L}(\vec{k})^2)((i \Omega -i k_n )^2 - \xi_{\nu}(\vec{k}-\vec{q})^2)} .
\end{align*}

With the analytic continuation ($i \Omega \rightarrow \xi_{\nu}(\vec{q}) + i\eta^+$), we obtain the imaginary part of the self-energy, 
\begin{align*}
\text{Im}&(\Pi_{L\rightarrow \nu}) =\\
&  \int_k \;  \frac{g_e^2\xi_{\nu}(\vec{q})  }{12} \left(2  k_i k_aP_{ia}(\vec{q}) + 2k_a k_j P_{ia}(\vec{q})P_{ji}(\vec{q}-\vec{k})   + 2    k_j k_b        P_{jb}(\vec{q}-\vec{k})                                    \right)    \\
& \quad \times\left[ n_B\left(\xi_{L}(\vec{k})\right) \frac{\left(\xi_{\nu}(\vec{q}) - \xi_{L}(\vec{k})\right) ^2}{2 \xi_{L}(\vec{k})}  \delta\left((\xi_{L}(\vec{k}) - \xi_{\nu}(\vec{q}))^2 - \xi_{\nu}(\vec{k}-\vec{q})^2         \right)\right. \\
&\quad\quad\quad \left.  + e^{{\beta\xi_{L}(\vec{k}) }} n_B\left(\xi_{L}(\vec{k})\right) \frac{\left(\xi_{\nu}(\vec{q})+\xi_{L}(\vec{k})\right) ^2}{2 \xi_{L}(\vec{k})}  \delta\left((\xi_{L}(\vec{k}) + \xi_{\nu}(\vec{q}))^2 - \xi_{\nu}(\vec{k}-\vec{q})^2         \right)  \right.\\
&\quad\quad\quad\quad\quad\left.+ n_B\left(\xi_{\nu}(\vec{k}-\vec{q})\right) \frac{\xi_{\nu}(\vec{k}-\vec{q})}{2}\delta\left( (\xi_{\nu}(\vec{q}) + \xi_{\nu}(\vec{k}-\vec{q}))^2 - \xi_{L}(\vec{k}) ^2\right) \right. \\
&\quad\quad\quad\quad\quad \left. + e^{\beta \xi_{\nu}(\vec{k}-\vec{q})} n_B\left( \xi_{\nu}(\vec{k}-\vec{q})\right)  \frac{\xi_{\nu}(\vec{k}-\vec{q})}{2} \delta\left( (\xi_{\nu}(\vec{q})- \xi_{\nu}(\vec{k}-\vec{q}))^2 - \xi_{L}(\vec{k})^2\right) \right]\\
&+ \int_k\; \frac{g_b^2 c^4}{12}  \left((\vec{k}\times\vec{q})^2(2q^2+ 2(\vec{q}-\vec{k})^2)    + 2 (\vec{q}\times(\vec{k}\times\vec{q}))\cdot ((\vec{q}-\vec{k})\times (\vec{k}\times\vec{q}))           \right) \frac{1} {4\xi_{\nu}(\vec{q}) \xi_{L}(\vec{k}) \xi_{\nu}(\vec{k}-\vec{q})} \nonumber  \\
&\quad\quad\times   \left[ \left(n_B\left(\xi_{L}(\vec{k}) \right)- n_B\left(-\xi_{\nu}(\vec{k}-\vec{q})\right)\right)\delta\left(\xi_{L}(\vec{k}) +\xi_{\nu}(\vec{k}-\vec{q}) - \xi_{\nu}(\vec{q})\right) \right.\\
& \quad \quad \quad \quad \left.- \left(n_B\left(- \xi_{L}(\vec{k}) \right)- n_B\left(-\xi_{\nu}(\vec{k}-\vec{q})\right)\right)\delta\left(-\xi_{L}(\vec{k}) +\xi_{\nu}(\vec{k}-\vec{q}) - \xi_{\nu}(\vec{q})\right)\right.\nonumber  \\
&\quad\quad\quad\quad\left.-\left(n_B\left(\xi_{L}(\vec{k}) \right)- n_B\left(\xi_{\nu}(\vec{k}-\vec{q})\right)\right)\delta\left(\xi_{L}(\vec{k}) -\xi_{\nu}(\vec{k}-\vec{q}) - \xi_{\nu}(\vec{q})\right) \right.\\
& \left.+\left(n_B\left(-\xi_{L}(\vec{k}) \right)- n_B\left(\xi_{\nu}(\vec{k}-\vec{q})\right)\right)\delta\left(-\xi_{L}(\vec{k}) -\xi_{\nu}(\vec{k}-\vec{q}) - \xi_{\nu}(\vec{q})\right) \right]\nonumber \\
&=  \frac{g_e^2 c}{48} \left(\frac{T}{c}\right)^5  f_1(\vec{y}) + \frac{g_b^2 c}{48} \left(\frac{T}{c}\right)^5  f_2(\vec{y})  = c g_e ^2 \left(\frac{T}{c}\right)^5 f_\nu\left(\frac{\xi_\nu(\vec{q})}{T},\alpha;g_b,g_e\right),
\end{align*}
where $ \vec{x}= \frac{c \vec{k}}{T}$, $\vec{y}= \frac{c \vec{q}}{T}$  and 
\begin{align*}
&f_1(\vec{y}) = \int_x \;y \left(2 x_i x_a P_{ia}(\vec{y})+2 x_a x_j P_{ia}(\vec{y}) P_{ji}(\vec{y}-\vec{x}) + 2 x_j x_b P_{jb}(\vec{y}-\vec{x})\right)(n(\alpha x)+n(\alpha x -y)) \frac{\alpha x -y}{{\alpha }x} \delta(\alpha x - |\vec{x}-\vec{y}|-\vec{y}), \\
&f_2(\vec{y}) = \int_x \left((\vec{x}\times \vec{y})^2(2 x^2+ 2(\alpha x -y)^2)+2(\vec{y}\times(\vec{x}\times\vec{y}))\cdot((\vec{y}-\vec{x})\times(\vec{x}\times\vec{y}))\right) \frac{n_B(\alpha x -y)-n_B(\alpha x)}{{\alpha} xy (\alpha x -y)}\delta(\alpha x - |\vec{x}-\vec{y}|-\vec{y}) ,
\end{align*}
and leading behaviors are in large $y$ and small  $y$ limits respectively,
\begin{align*}
&f_1(y\ll1)\simeq \frac{192\pi y^4}{\alpha^5} , \quad f_2(y \ll1 ) \simeq \frac{256\pi y^4}{3\alpha ^5} ,\\
&f_1(y\gg1)\simeq  \frac{128 \pi y^5}{\alpha^4}  e^{-y}, \quad f_2(y\gg1) \simeq \frac{ 512 \pi y^5}{3\alpha^4}  e^{-y}.
\end{align*}
The overall profile of the functions ($f_1,f_2$) are illustrated in Fig. A1.

\begin{figure}[h!]
\centering
\includegraphics[width=0.45\textwidth]{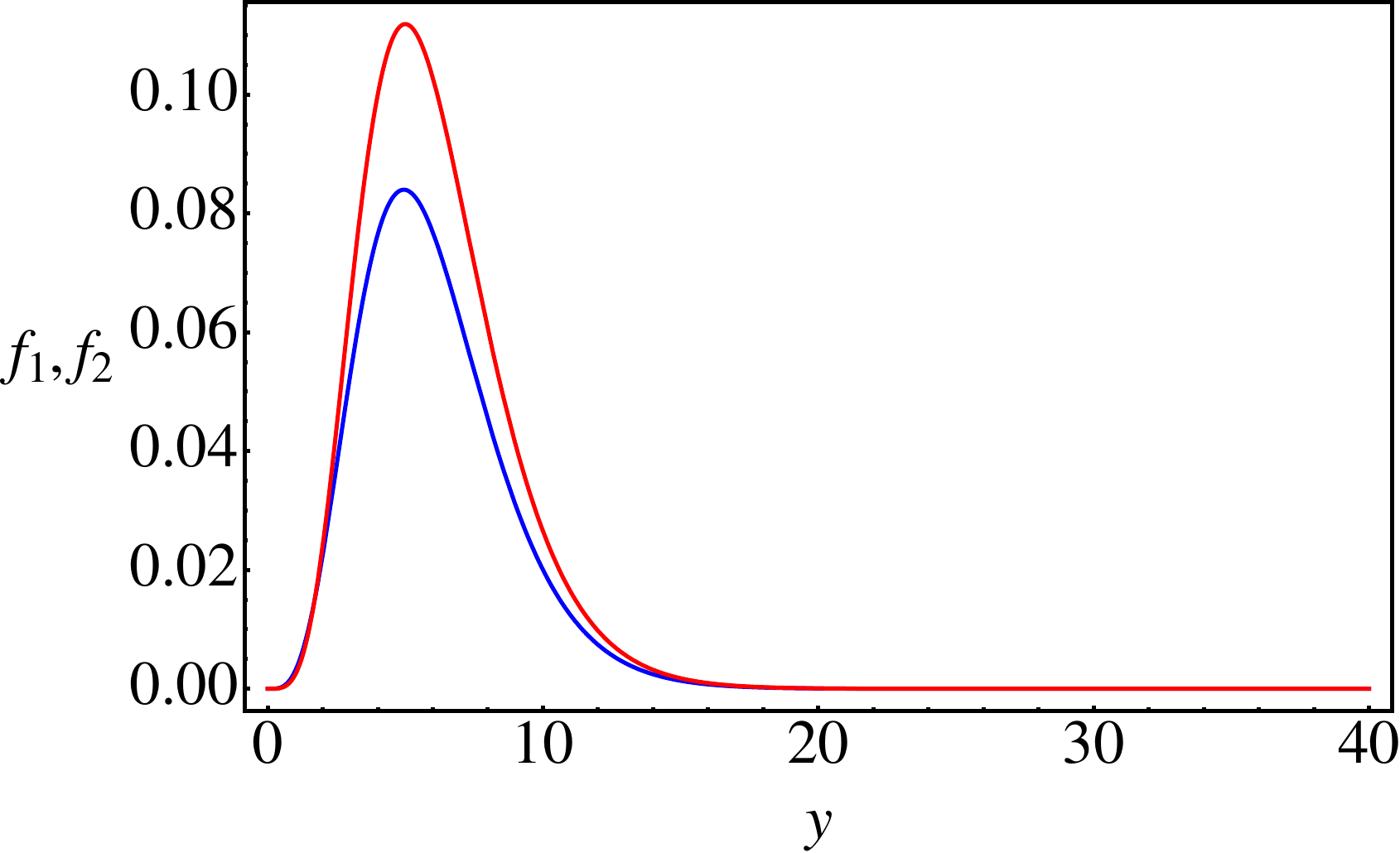}
\caption{$f_1(\vec{y})$ and $f_2(\vec{y})$ with $\alpha = v/c  =10$. The blue (red) line is for  $f_1$ ($f_2$). }
\end{figure}

{

{Photons also scatter with E-particles. Interaction Hamiltonian is $H_{\nu-E}\left(\left(\vec{p}-\frac{e}{c}\vec{a}\right)^2,...,\Delta_E\right)$ where $\vec{p}$, $e$, $\Delta_E$ is momentum, gauge charge and mass of E-particle respectively. In counting of photon self-energy from E-particle in a diagramatic way, diagram must contain even number($2n$) of E-particle propagator generally. Since E-particle mass is the largest scale in $H_{\nu-E}$ and temperature is much lower than this mass scale, one can see that every diagram containing $2n$ of E-particle propagator is proportional to $e^{-n\frac{\Delta_E}{T}}$ regardless of dispersion relation of E-particle. For more quantitative calculation, we assume the simplest quadratic Hamiltonian of E-particle,}
\begin{align*}
H_{\nu -E}  =\Delta_E c^2 + \frac{1}{2\Delta_E} (\vec{p}- \frac{e}{c} \vec{a})^2,
\end{align*}
with E-particle current, $\vec{j}_E   = \frac{-i e}{2\Delta_E c }  (\varphi^* \vec{\nabla} \varphi - \varphi \vec{\nabla} \varphi^*) $. $\Delta_E$  and $e$ are mass and gauge charge of E-particle. Photon self-energy from E-particle is,
\begin{align*}
&\begin{tikzpicture}
\draw[vector,thick](-2,0)--(-1,0);
\draw[vector,thick](1,0)--(2,0);
\draw[fermion,thick](-1,0)arc(180:0:1);
\draw[fermionbar,thick](1,0)arc(0:-180:1);
\end{tikzpicture}
\end{align*}
\begin{align*} 
&\rightarrow \Pi_{E\rightarrow \nu}(\vec{q})  =T \sum_{p_n}\frac{1}{i \Omega}  \left(\frac{e}{2\Delta_Ec} \right)^2 \int_p  (2\vec{p}+\vec{q})^2  \frac{1}{i p_n + i \Omega -\xi_E (\vec{p}+\vec{q}) } \frac{1}{ ip_n  - \xi_E (p)} ,
\end{align*}
where $\vec{q}$ is photon momentum and $\xi_e(\vec{p}) = \Delta_E c ^2 + \frac{p^2}{2\Delta_E}$. 
In the same process with phonon case, imaginary part of the self-energy is
\begin{align*}
\text{Im}(\Pi_{E\rightarrow \nu})=e^{-\frac{\Delta_Ec ^2}{T}} (1- e^{-\frac{ c q      }{ T}})\frac{\Delta_E}{c q^2}\left(\frac{e}{2\Delta_Ec}\right)^2 2\pi \int_{\left|\Delta_E (c-\frac{q}{2\Delta_E})\right|}^{\Delta_E c}   dp\;\; p (4 p^2 + 4 \Delta_E c q     -q^2) e^{-\frac{p^2}{2\Delta_E T }}  .
\end{align*}
In the last equality, cut-off is introduced to regularize divergence coming from artificial dispersion relation which is valid in small momentum limit. One can see that life-time corrected by E-particle proportional to $e^{-\frac{\Delta_{E}}{T}}$ in the unit of $c =1$.

}
\subsubsection{Phonon decay rate}
The phonon self-energy from photon is 
\begin{align*}
&\begin{tikzpicture}
\draw[dotted,thick](-1.5,0)--(-.9,0);
\draw[vector,thick](-.9,0)arc(180:0:1);
\draw[vector,thick](-.9,0)--(1.1,0);
\draw[dotted,thick](1.1,0)--(1.5,0);
\end{tikzpicture}  \nonumber 
\end{align*}
\begin{align*}
&\rightarrow \Pi _{ \nu \rightarrow L} = T \sum_{q_n}\int_q \frac{I_2 } {((i q_n)^2 - \xi_{\nu}(\vec{q})^2)(( i\Omega - i q_n )^2 - \xi_{\nu}(\vec{k}-\vec{q})^2)} ,
\end{align*}
where 
\begin{align*}
12 I_2  =&  \frac{1}{i \Omega}\left( (k_i \epsilon_{L j }+ k_j \epsilon_{L i} ) g_e (- i q_n ) \epsilon_{\nu q i}^{*} (-i \Omega+ i q_n ) \epsilon_{\nu k-q j }^{ *} (-k_a \epsilon_{L b} ^{*}- k_b \epsilon_{L a } ) g_e (i q_n) \epsilon_{\nu q a} (i \Omega- i q_n ) \epsilon_{\nu k-q b }  \right. \\
&\left. + (k_i \epsilon_{L j} + k_j \epsilon_{L i} ) g_b c ^2   (-\vec{q}\times \vec{\epsilon}_{\nu q}^* )_i (-(\vec{k}-\vec{q})\times \vec{\epsilon}_{\nu k-q }^*)_j  (-k_a \epsilon_{L b} ^{*}- k_b \epsilon_{L a} ^{*}) g_b c ^2   (\vec{q}\times \vec{\epsilon}_{\nu q })_a ((\vec{k}-\vec{q})\times \vec{\epsilon}_{\nu k-q })_b\right).
\end{align*}
One can do similar calculations as in the photon self-energy,  
\begin{align*}
&\Pi_{\nu \rightarrow L} =  \frac{1}{i\Omega} \int _q \left(\frac{g_e ^2}{12} \left(2k_i k_a P_{ia}(\vec{q}) +2k_a k_j P_{ia}(\vec{q})P_{ji}(\vec{q}-\vec{k}) + 2 k_j k_b P_{jb}(\vec{q}-\vec{k})                    \right)   T \sum_{q_n}  \frac{-(iq_n)^2 (i\Omega- iq_n)^2}{\left((iq_n)^2 - \xi_{\nu}(\vec{q})^2\right) ((i \Omega - iq_n)^2 -\xi_{\nu}(\vec{k}-\vec{q})^2)} \right. \\
&\left. +\frac{g_b^2 c ^4}{12}  \left( 2  ((\vec{k}-\vec{q})^2+q^2) (\vec{k}\times \vec{q})^2  + 2 ((\vec{k}-\vec{q}) \times (\vec{k}\times \vec{q}))\cdot (\vec{q} \times (\vec{k}\times\vec{q}))\right)            T \sum_{q_n} \frac{-1}{((i q_n)^2 - \xi_{\nu}(\vec{q})^2 )((i\Omega -iq_n)^2 - \xi_{\nu}(\vec{k}-\vec{q})^2)} \right),
\end{align*}
and the imaginary part is  
\begin{align*}
\text{Im}(\Pi_{\nu \rightarrow L }) =& \int_q \; \frac{g_e^2}{12\xi_{L} (\vec{k})}        (2k_i k_a P_{ia}(\vec{q}) +2k_a k_j P_{ia}(\vec{q})P_{ji}(\vec{q}-\vec{k}) + 2 k_j k_b P_{jb}(\vec{q}-\vec{k})                    )       \\
&\quad\times\left[n_B\left(\xi_{\nu}(\vec{q})\right)\frac{\xi_{\nu}(\vec{q})}{2 }  \left(\xi_{\nu}(\vec{q})- \xi_{L}(\vec{k}) \right)^2{\delta\left((\xi_{\nu}(\vec{q})- \xi_{L}(\vec{k}) )^2- \xi_{\nu}(\vec{k}-\vec{q})^2\right)} \right.\\
& \quad \quad \left.  + e^{\beta \xi_{\nu}(\vec{q})} n_B\left(\xi_{\nu}(\vec{q})\right) \frac{ \xi_{\nu}(\vec{q})}{2 } \left(\xi_{\nu}(\vec{q}) + \xi_{L}(\vec{k})\right)^2 \delta\left({(\xi_{\nu}(\vec{q})+ \xi_{L}(\vec{k}) )^2 - \xi_{\nu}(\vec{k}-\vec{q})^2}\right)\right.\\
&\quad\quad\quad\left.  + n_B\left(\xi_{\nu}(\vec{k}-\vec{q})\right) \frac{\xi_{\nu}(\vec{k}-\vec{q})}{2} {\left(\xi_{L}(\vec{k})+\xi_{\nu}(\vec{k}-\vec{q})\right)^2}\delta\left((\xi_{L}(\vec{k})+\xi_{\nu}(\vec{k}-\vec{q}))^2 - \xi_{\nu}(\vec{q})^2\right)\right.  \\
&\quad\quad\quad\left.+ e^{\beta \xi_{\nu}(\vec{k}-\vec{q})} n_B\left(\xi_{\nu}(\vec{k}-\vec{q})\right)  \frac{\xi_{\nu}(\vec{k}-\vec{q})}{2} {\left(\xi_{L}(\vec{k})-\xi_{\nu}(\vec{k}-\vec{q})\right)^2 }\delta\left((\xi_{L}(\vec{k})-\xi_{\nu}(\vec{k}-\vec{q}))^2 - \xi_{\nu}(\vec{q})^2\right)\right]\\
&+ \int_q \; \frac{g_b^2 c ^4}{12 }       \left( 2  ((\vec{k}-\vec{q})^2+q^2) (\vec{k}\times \vec{q})^2  + 2 ((\vec{k}-\vec{q}) \times (\vec{k}\times \vec{q}))\cdot (\vec{q} \times (\vec{k}\times\vec{q}))\right)         \frac{1} {4 \xi_{\nu}(\vec{q}) \xi_{L}(\vec{k}) \xi_{\nu}(\vec{k}-\vec{q})}             \\
&\quad\quad\times   \left[ \left(n_B\left(\xi_{\nu}(\vec{q}) \right)- n_B\left(-\xi_{\nu}(\vec{k}-\vec{q})\right)\right)\delta\left(\xi_{\nu}(\vec{q})+\xi_{\nu}(\vec{k}-\vec{q}) - \xi_{L}(\vec{k})\right)\right.\\
&\quad \quad \quad \left. - \left(n_B( \xi_{\nu}(\vec{q}))- n_B\left(\xi_{\nu}(\vec{k}-\vec{q})\right)\right)\delta\left(\xi_{\nu}(\vec{q})-\xi_{\nu}(\vec{k}-\vec{q}) - \xi_{L}(\vec{k})\right)\right. \\
&\quad\quad\quad\quad\left.-\left(n_B\left(-\xi_{\nu}(\vec{q})\right)- n_B\left(-\xi_{\nu}(\vec{k}-\vec{q})\right)\right)\delta\left(-\xi_{\nu}(\vec{q})+\xi_{\nu}(\vec{k}-\vec{q}) - \xi_{L}(\vec{k})\right) \right.\\
&\left. +\left(n_B\left(-\xi_{\nu}(\vec{q})\right)- n_B\left(\xi_{\nu}(\vec{k}-\vec{q})\right)\right)\delta\left(-\xi_{\nu}(\vec{q})-\xi_{\nu}(\vec{k}-\vec{q}) - \xi_{L}(\vec{k}) \right)\right] \\
&=  \frac{g_e^2 c}{48} \left(\frac{T}{c}\right)^5  f_3(\vec{x}) + \frac{g_b^2 c}{48} \left(\frac{T}{c}\right)^5  f_4(\vec{x}) = c g_e^2 \left(\frac{T}{c}\right)^5 f_L \left( \frac{\xi_L(\vec{k})}{T},\alpha;g_b,g_e\right) ,
\end{align*}
where
\begin{align*}
&f_3(\vec{x}) = \int_y \left(2 x_i x_a P_{ia}(\vec{y})+2 x_a x_j P_{ia}(\vec{y}) P_{ji}(\vec{y}-\vec{x}) + 2 x_j x_b P_{jb}(\vec{y}-\vec{x})\right)\frac{y (\alpha x -y)}{\alpha x}  (n_B(y)+1+  n_B(\alpha x -y))  \delta(\alpha x - |\vec{x}-\vec{y}|-\vec{y}), \\
&f_4(\vec{x}) = \int_y \left(( 2(\alpha x -y)^2+2 y^2)(\vec{x}\times \vec{y})^2+2((\vec{y}-\vec{x})\times(\vec{x}\times\vec{y}))   \cdot  (\vec{y}\times(\vec{x}\times\vec{y}))   \right) \frac{n_B(y)-n_B(-(\alpha x -y))}{\alpha xy(\alpha x -y)}\delta(\alpha x - |\vec{x}-\vec{y}|-\vec{y}) ,
\end{align*}
and asymptotic behavior is,
\begin{align*}
& f_3(x\ll1 ) \simeq 2\pi x^4 \alpha ^2, \quad f_4(x\ll 1 ) \simeq \frac{8}{5}\pi x^4 \alpha ^2 ,\\
& f_3(x\gg1 )\simeq \frac{1}{2} \pi x^5 \alpha ^3, \quad f_4(x\gg 1) \simeq \frac{2}{3} \pi x^5 \alpha ^3.
\end{align*}
{ For the sound attenuation experiments,  interesting low frequency limit ($\xi_L=\omega \ll T$) behaviors of $ \frac{1}{\tau_{L,tot}(\omega;B_{ext}=0)} -\frac{1}{\tau_{L,tot}(\omega;B_{ext}\rightarrow\infty)} $ in the main-text can be understood by $x\ll 1$ limit behaviors of $f_3$, $f_4$. Simple ways to understand this results as follows. Once scaling out photon and phonon momentums by input frequency $(\omega)$, the power counting of phonon self-energy calculations from photons gives $\omega^5$ from matrix element and integral measure and distribution factor gives $\frac{T}{\omega}$.}

The overall profile of the functions ($f_3,f_4$) are illustrated in Fig. A2.

\begin{figure}[h!]
\centering
\includegraphics[width=0.45\textwidth]{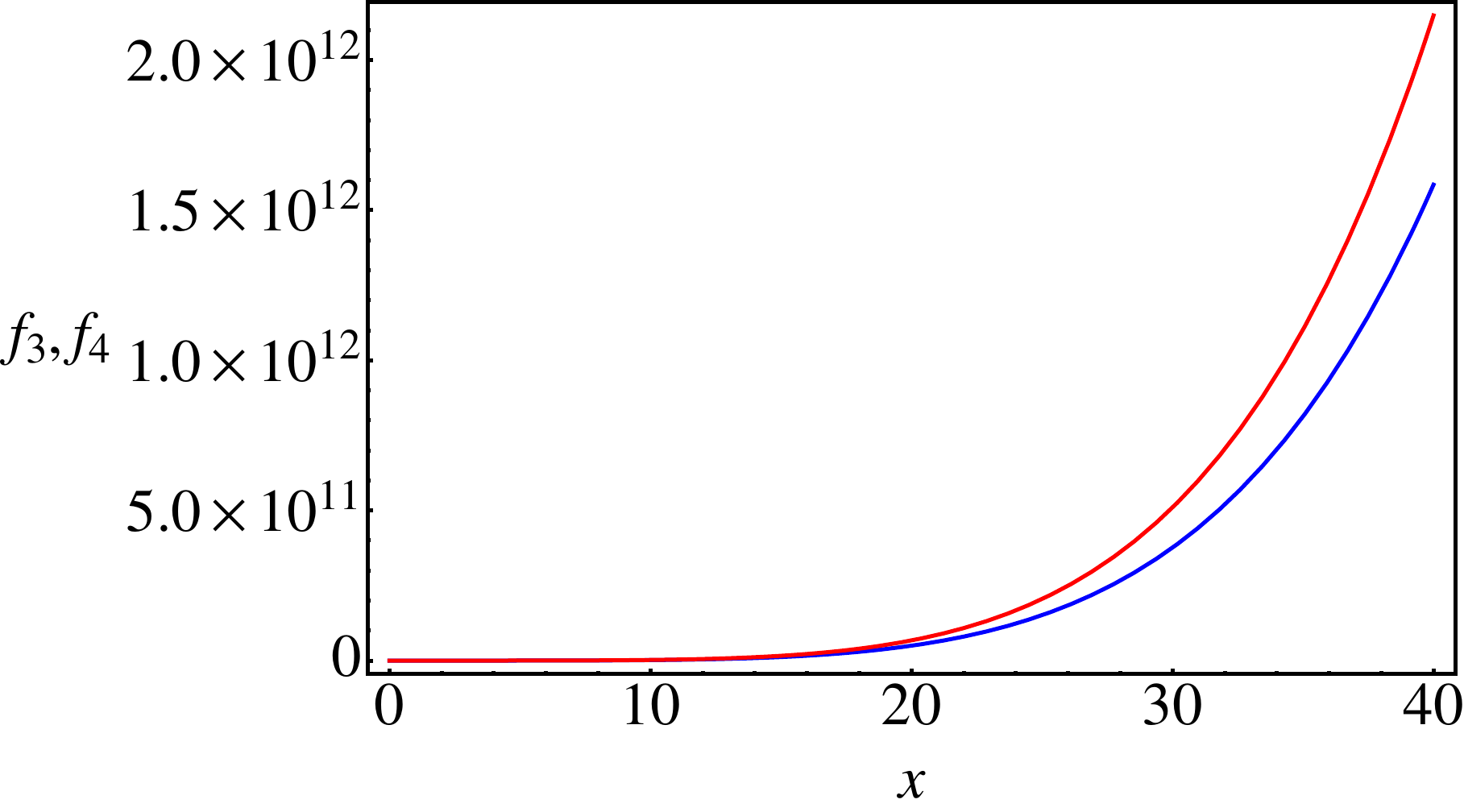}
\caption{$f_3(\vec{x})$ and $f_4(\vec{x})$ with $\alpha = v/c  =10$. The blue (red) line is for  $f_3$ ($f_4$).}
\end{figure}

{
Along the same line with photon case, phonon life-time is also corrected by E-particle with the Hamiltonian,
\begin{align*}
H_{L-e} =\int_x  \mathcal{P}_{\alpha \beta } [ j_{ e}^\alpha j_{ e}^\beta +\partial^\alpha \rho_e \partial^\beta \rho_e  ],
\end{align*}
and we are going to ignore the second term because it contains more derivative. Leading contribution diagram is,
\begin{align*}
&\begin{tikzpicture}
\draw[dotted,thick](-1.7,0)--(-1,0);
\draw[dotted,thick](1.7,0)--(1,0);
\draw[fermion,thick](-1,0)arc(180:0:1);
\draw[fermion,thick](1,0)arc(360:180:1);
\draw[fermion,thick](1,0)arc(40:140:1.3);
\draw[fermion,thick](-1,0)arc(-140:-40:1.3);
\end{tikzpicture}\nonumber  
\end{align*}
\begin{align*}
\rightarrow &\Pi _{E\rightarrow \nu} = \\
&\frac{4\cdot 3}{2} \frac{g_{LE}^2}{i\Omega}  \frac{1}{2}\sum_{pol} \int_{p_1,p_2,p_3} \left(\frac{e}{2\Delta_E c} \right)^4 ( k_a \epsilon_{Lb} + k_b \epsilon_{L a})  ( p_{1 a} + p_{2 a}) ( p_{3 b} + p_{4 b} )( k_c \epsilon_{Ld}^* + k_d \epsilon_{L c}^*)  ( p_{1 c} + p_{2 c}) ( p_{3 d} + p_{4 d} ) \\
&\quad\quad \times T^3 \sum_{p_n,p_l,p_m} \frac{1}{i p_n -\xi_e(\vec{p}_1)}\frac{1}{i p_l -\xi_e (\vec{p}_2)}\frac{1}{i p_m -\xi_e (\vec{p}_2)} \frac{1}{i p_n - i p_l + i p_m -i\Omega - \xi_e (\vec{p}_4)}, \\
\end{align*}
where $\vec{k}$ is phonon momentum and $\vec{p}_4 = \vec{p}_1 -\vec{p}_2+\vec{p}_3 -\vec{k}$, $\frac{4\cdot 3}{2}$ is from combinatorics. Imaginary part  of the self-energy is, 
\begin{align*}
\text{Im}(\Pi_{E\rightarrow \nu})&=\\
&-6 \frac{g_{LE}^2}{i\Omega} \frac{1}{2}\sum_{pol} \int_{p_1,p_2,p_3} \left(\frac{e}{2\Delta_E c} \right)^4 ( k_a \epsilon_{Lb} + k_b \epsilon_{L a})  ( p_{1 a} + p_{2 a}) ( p_{3 b} + p_{4 b} )( k_c \epsilon_{Ld}^* + k_d \epsilon_{L c}^*)  ( p_{1 c} + p_{2 c}) ( p_{3 d} + p_{4 d} )\\
&\quad \times\left(n_B(\xi_e(\vec{p}_1))- n_B(\xi_e(\vec{p}_4))\right)(n_B(\xi_e(\vec{p}_2))- n_B(-\xi_e(\vec{p}_4)+\xi_e(\vec{p}_1)               ))(n_B(\xi_e(\vec{p}_3))- n_B\left(\xi_e(\vec{p}_4)-\xi_e(\vec{p}_1)+\xi_e(\vec{p}_2)                )\right)\\
&\quad \quad \quad \quad \delta\left(\xi_L(\vec{k}) -\xi_e(\vec{p}_1)+\xi_e(\vec{p}_2)-\xi_e(\vec{p}_3)+\xi_e (\vec{p}_4)\right).
\end{align*}

If $T\ll \Delta_E c ^2$,
\begin{align*}
&n_B(\xi_e(\vec{p}_1) ) - n_B(\xi_e(\vec{p}_4)) \simeq e^{-\frac{\Delta_Ec^2}{T}} (e^{-\frac{p_1^2}{2\Delta_ET}}-e^{-\frac{p_4^2}{2\Delta_ET}}),\nonumber\\
&n_B(\xi_e (\vec{p}_3) ) - n_B(\xi_e (\vec{p}_4)-\xi_e{(\vec{p}_1)}+\xi_e {(\vec{p}_2)}) \simeq e^{-\frac{\Delta_E c^2}{T}} (e^{-\frac{p_3^2}{2\Delta_ET}}-e^{-\frac{p_4^2}{2\Delta_ET} +\frac{p_1^2}{2\Delta_ET}-\frac{p_2^2}{2\Delta_ET} }),\nonumber\\
&n_B(\xi_e{(\vec{p}_2)} ) - n_B(-\xi_e{(\vec{p}_4) }+\xi_e{(\vec{p}_1)}) \simeq e^{-\frac{\Delta_Ec^2 }{T}} e^{-\frac{p_1^2}{2\Delta_ET}} - \frac{1}{e^{{\frac{p_1^2}{2\Delta_E T} - \frac{p_4^2}{2\Delta_E T} }-1}}.
\end{align*}
Therefore,
\begin{align*}
\text{Im}({\Pi_{E\rightarrow \nu}}) \simeq 6\frac{e^4 g_{LE}^2}{v^2 c^{10}}  \left(\frac{2\Delta_E c^2}{T}\right)^{\frac{5}{2}} \frac{T^{10}}{k}  e^{-2\frac{\Delta_E c^2}{T}} f(\vec{K},\frac{\Delta_E v^2}{T}),
\end{align*}
where
\begin{align*}
f(\vec{K},\frac{\Delta_E v^2 }{T}) =& \frac{1}{2}\sum_{pol} \int_{P_1,P_2,P_3} ( K_a \epsilon_{Lb} + K_b \epsilon_{L a})  ( P_{1 a} + P_{2 a}) ( P_{3 b} + P_{4 b} )( K_c \epsilon_{Ld}^* + K_d \epsilon_{L c}^*)  ( P_{1 c} + P_{2 c}) ( P_{3 d} + P_{4 d} )\\
&\quad  \times (e^{-P_1^2} -e^{-P_4^2} ) (e^{-P_3^2} -e^{-P_4^2+P_1^2-P_2^2} ) \frac{1}{e^{P_1^2-P_4^2}-1}    \delta(K - P_1^2 + P_2^2 -P_3^2 +P_4^4) ,
\end{align*}
with $\vec{K} = \frac{v \vec{k} }{T}$, $\vec{P}_{i=1,2,3} =\frac{\vec{p}_i }{\sqrt{2\Delta_E T}}$, $\vec{P}_4 = \vec{P}_1-\vec{P}_2 +\vec{P}_3-\sqrt{\frac{T}{2\Delta_E v^2}}\vec{K}$. In the last equality, we ignored the first term in $n_B(\xi_e(\vec{p}_2))-n_B(\xi_e(\vec{p}_2))$.  Phonon life-time correction is proportional to $e^{-2\frac{\Delta_E}{T}}$ in the unit of $c=1$.
}

\begin{center} 
\textbf{Appendix B: Boltzmann equation }
\end{center}

We employ the Boltzmann transport for thermal conductivity calculation.
The thermal conductivity is related to the entropy production,
\begin{align*}
\frac{1}{\kappa_{tot}} \frac{\vec{U}^2_{tot}}{T} = \dot {s}_{scatt} = \dot {s}_{L-\nu}+\dot {s}_{L-boundary}+\dot {s}_{\nu-boundary}.
\end{align*}
The total heat current is $\vec{U}_{tot} = \vec{U}_{L} +\vec{U}_{\nu}$, and $\dot{s}_{p-q}$ is the entropy production rate with a scattering between p and q degrees of freedom. Recall that we ignore E, M particles by considering $T\ll \Delta_{E,M}$ and assume clean enough systems ignoring impurity scatterings.

At low enough temperature , all scattering except boundary ones are suppressed scattering mean free paths to be a system size. Then, the thermal conductivity is determined by specific heat and velocity, which gives,
\begin{align*}
\frac{\kappa_{\nu}}{\kappa_{tot}} = \frac{c_\nu c}{c_\nu c + c_L v } \rightarrow \frac{v ^2 }{c^2 + v ^2 }  = \frac{\alpha ^2 }{\alpha ^2 + 1} .
\end{align*}
Thus, the thermal conductivity is dominated by the photon in the limit of $\alpha \gg 1$. 

Increasing temperatures, scattering from $H_{L-\nu}$ becomes important. We introduce a conventional variational ansatz,  $ n_{L(\nu)} = n^0 _{L(\nu)} -\Phi_{L(\nu)} \frac{\partial n^0 _{L(\nu)}}{\partial \xi}$ where $n^0_{L(\nu)}$ is an equilibrium phonon (photon) distribution function. After linearize the Boltzmann equation, we minimize the thermal resistivity, $\frac{1}{\kappa_{tot}}$, 

\begin{align*}
\frac{1}{\kappa_{tot}} = \frac{ \frac{1}{2T^2} \int_{q,q',k}\;  \left(\Phi_{\nu}(\vec{q}) +\Phi_{\nu}(\vec{q'})-\Phi_{L}(\vec{k})\right)^2 \mathcal{P}_{\nu \nu \rightarrow L} \; \delta^{(3)} (\vec{q}+\vec{q'} - \vec{k}) \delta(\xi_{L}(\vec{k})  - \xi_{\nu}(\vec{q})- \xi_{\nu}(\vec{ q'})) }{\left|\int_q\;  \vec{v}_{\nu} \Phi_\nu \frac{\partial n_{\nu}^0}{\partial T}+\int _k\;  \vec{v}_{L} \Phi_L \frac{\partial n_{L}^0}{\partial T}\right|^2},
\end{align*}
where
\begin{align*}
&\mathcal{P}_{\nu \nu \rightarrow L}= \frac{1}{3\cdot 4}\sum_{pol} (1+n_{\nu }^0(q)){(1+ n_{\nu }^0(q'))}n_{L}^0(k)\left| \mathcal{A}_{\nu\nu\rightarrow L} \right|^2  .
\end{align*}
The transition rate of the scattering process $|\mathcal{A}_{\nu\nu\rightarrow L}|^2$ is introduced. With $H_{L-\nu}$, the transition rate can be straightforwardly obtained by the Fermi-golden rule. 
Averaging over the polarizations, we obtain the transition rate,
\begin{align*}
\frac{1}{12} \sum_{pol} \left|\mathcal{A}_{\nu\nu\rightarrow L}\right|^2 =\frac{2\pi}{12} \frac{ g_e^2 \xi_{\nu}(\vec{q})^2 \omega_{\nu q'}^2 \left(2(q^2 +q'^2)+3\vec{q}\cdot\vec{q'} + \frac{(\vec{q}\cdot\vec{q'})^3}{q^2 q'^2}\right)+2 g_b ^2 c ^4    (\vec{q}\times\vec{q'})^2 (\vec{q}-\vec{q'})^2 }  {(2\pi)^5 2^3 \xi_{\nu}(\vec{q}) \xi_{\nu}(\vec{ q'}) \xi_{L}(\vec{k})}.
\end{align*}

We use the ansatz, $\Phi_{\nu,L}(\vec{q}) = (\hat{u}\cdot\vec{q}) \sum_{m=1} ^r a_{\nu,L,m} q^{2m} $. Notice that the ansatz captures normal scattering processes, which is reasonable for low temperature thermal transport. Also, by symmetry, another type of term, $(\hat{u}\cdot \vec{q})^{2m}$, is allowed, but we numerically check this term is sub-dominant.

Setting  $\frac{c \vec{q}}{T} = \vec{Q}$, $\frac{c \vec{q'}}{T}=\vec{Q'}$, $\frac{c \vec{k}}{T}=\vec{K}$, we have the final formula to minimize, 
\begin{align*}
\frac{1}{\kappa_{tot}} = \frac{ T^2 } {(2\pi)^4 96c^3  }  \frac{\int _{Q ,Q',K } (1+n_\nu^0 ({Q}))(1+n_\nu^0 ({Q'})) n_L ^0(\alpha {K})(\tilde{\Phi}_\nu(\vec{Q}) + \tilde{\Phi}_\nu({\vec{Q'}}) - \tilde{\Phi}_L({\vec{K}})    )^2  \mathcal{\tilde{A}}_{\nu\nu\rightarrow L} } { \left|\int_Q \tilde{\Phi}_\nu (\vec{Q})\vec{Q}\frac{\partial n_\nu ^0 ({Q})}{\partial Q}+ {\int_K  \tilde{\Phi}_L (\vec{K})\alpha \vec{K}\frac{\partial n_L ^0 (\alpha { K})}{\partial K}} \right|^2},
\end{align*}

with
\begin{align*}
&\mathcal{\tilde{A}}_{\nu\nu\rightarrow L} =  \frac{ g_e^2 Q^2 Q'^2  (2 (Q^2+Q'^2 )+3\vec{Q}\cdot\vec{Q'}+  \frac{(\vec{Q}\cdot\vec{Q'})^3}{Q^2Q'^2}) + 2g_b ^2 (\vec{Q}\times\vec{Q'})^2 (\vec{Q}-\vec{Q'})^2}{\alpha QQ'K}\delta^{(3)}(\vec{Q}+\vec{Q'} - \vec{K})  \delta(Q+Q'-\alpha K),\\
&\tilde{\Phi}_{\nu, L}(\vec{Q})=  (\hat{u}\cdot\vec{Q}) \sum_{m=0}^r \tilde{a}_{m({\nu, L})} Q^{2m}. 
\end{align*}
Remark that the temperature dependence is $\kappa_{tot} \propto \frac{1}{T^2}$, which can be also obtained by scaling analysis. The temperature dependence can be understood by counting dimensions of specific heat $(c_{\nu, L}\propto T^3 )$ and the decay rate ($\tau^{-1} \propto T^5$).

and we find that the case with the phonons in equilibrium has much smaller thermal resisvitiym which demonstrates that the thermal conductivirt is mainly carried by emergent photons.

To determine main carriers, we use two sets of ansatz,  1) phonons in equilibrium ($\Phi_L=0$) and 2) photons in equilibrium ($\Phi_\nu=0$).
Our numerical calculations up to the order $r$ are as follows,  
\begin{align*}
\begin{tabular}{l | {c}|{c}r} 
1/$\kappa_{tot}$            &  $\Phi_L=0$ &  $\Phi_\nu=0$ \\
\hline
$r=0$  & $9.791\times 10^{-5}$ & $9.791\times 10$\\
$r=1$   & $8.833 \times 10^{-5}$ & $4.359\times 10$ \\
$r=2$   & $7.594\times 10^{-5}$ & $2.784\times 10$\\
\end{tabular}
\end{align*}
and we conclude that the case with the phonons in equilibrium has much smaller thermal resistivity which demonstrates that the thermal conductivity is mainly determined by emergent photons.

\end{document}